\documentclass[aps, prm, twocolumn, amssymb, amsmath, float, superscriptaddress]{revtex4-1}
\usepackage{graphicx,dcolumn,bm,hyperref,subfigure,multirow,color}
\begin{document}

\title{Probabilistic deep learning approach for targeted hybrid organic-inorganic perovskites}

\author{Vu Ngoc Tuoc}
\affiliation{Institute of Engineering Physics, Hanoi University of Science and Technology, 1 Dai Co Viet Rd., Hanoi, Vietnam}
\author{Nga T. T. Nguyen}
\affiliation{CCS-3, Information Sciences, Los Alamos National Laboratory, Los Alamos, NM, 87545, USA}
\author{Vinit Sharma}
\affiliation{National Institute for Computational Sciences, Oak Ridge National Laboratory, Oak Ridge, TN 37831, USA}
\affiliation{Joint Institute for Computational Sciences, University of Tennessee, Knoxville, TN 37996, USA}
\author{Tran Doan Huan}
\email{huan.tran@mse.gatech.edu}
\affiliation{School of Materials Science and Engineering, Georgia Institute of Technology, 771 Ferst Dr. NW, Atlanta, GA 30332, USA}

\date{\today}

\begin{abstract}
We develop a probabilistic machine learning model and use it to screen for new hybrid organic-inorganic perovskites (HOIPs) with targeted electronic band gap. The data set used for this work is highly diverse, containing multiple atomic structures for each of 192 chemically distinct HOIP formulas. Therefore, any property prediction on a given formula must be associated with an irreducible “uncertainty” that comes from its unknown atomic details. As a result, dozens of new HOIP formulas with band gap falling between 1.25 and 1.50 eV were identified and validated against suitable first-principles computations. Through this demonstration we show that the probabilistic deep learning approach is robust, versatile, and can be used to properly quantify this uncertainty. In conclusion, the probabilistic standpoint and approach described herein could be widely useful for the very common and inevitable data uncertainty which is rooted at the incompleteness of information during experiments and/or computations.
\end{abstract}

\maketitle

\section{Introduction}\label{sec:intro}
The enormous interest devoted to hybrid organic-inorganic perovskites (HOIPs), specifically methylammonium lead iodide CH$_3$NH$_3$PbI$_3$ \cite{Kojima:perovskites}, during the last decade was mainly fueled by the amazing power conversion efficiency when they are used as solar absorbers \cite{Burschka:2013, Liu:perovskites, Hao2014, Saparov:ChemRev16, perovskite_20pc}. Materials in this family have the chemical formula of ABX$_3$ and adopt the classic perovskite structures in which cations A are inserted into the cages formed by the 3D network of cations B and anions X. Because the organic cations are highly anisotropic and typically much larger than any inorganic cation A of classic perovskites, the BX$_3$ network is inevitably deformed/broken, introducing remarkable structural diversity \cite{Mitzi95, Mitzi, Huan:perovskites, Chiho:hybrid, flores2018emergence, Saparov:ChemRev16}, and in some cases, enabling the suitability of HOIPs for other applications as well, i.e., optoelectronics \cite{long2020chiral,Minh:2dperov}, spintronics \cite{zhai2017giant}, and ferroelectrics \cite{hou2020two}.

The most notable HOIPs, i.e., methylammonium lead iodide CH$_3$NH$_3$PbI$_3$ and formamidinium lead iodide HC(NH$_2$)$_2$PbI$_3$, are essentially unstable and contain a toxic species (Pb). Therefore, searches for new HOIPs have been highly active \cite{lu2018accelerated, li2020robot, gebhardt2019mix,wu2019global, jacobs2019materials, tuoc2020lead}. While Sn is probably the most examined alternative for Pb, about a hundred mono-valence organic cations were screened at some levels of experiments and computations, especially using machine-learning (ML) techniques \cite{Saparov:ChemRev16, Huan:perovskites, Chiho:hybrid, li2020robot, jacobs2019materials, nguyen2021novel, gebhardt2019mix, wu2019global, yilmaz2020critical}. These works typically start by putting together a dataset of the ABX$_3$ formulas and some properties needed for a solar absorber. Then, some ML models were developed, directly mapping the HOIP formulas onto the properties. Next, a large-scale screening follows, utilizing these models to identify those having favorable properties for specific applications. Within this generic workflow, a representative atomic structure was assumed for each formula, using which some targeted properties, e.g., the electronic band gap $E_{\rm g}$, can be computed. The main reason to directly map the ABX$_3$ formula onto the respective property is that predicting reasonable atomic structures for a large number of formulas, e.g., using computations,\cite{OganovBook} is very expensive and technically impractical. In a vast majority of these ML works \cite{lu2018accelerated, li2019predictions, wu2019global, li2021high} the (3D) perovskite prototype structure, whose $E_{\rm g}$ often falls into the right window for solar cell applications, was selected as the representative for each formula. This is an oversimplification because multiple phases of a HOIP can co-exist at the same condition, e.g., the cubic and tetragonal phases of CH$_3$NH$_3$PbI$_3$ can be realized at the room temperature \cite{palazon2019room}. Therefore, the previously used datasets \cite{lu2018accelerated, li2019predictions, wu2019global, li2021high} are generally small (a few hundreds entries) and structurally uniform.

\begin{figure}[t]
\centering
\includegraphics[width=1.0\linewidth]{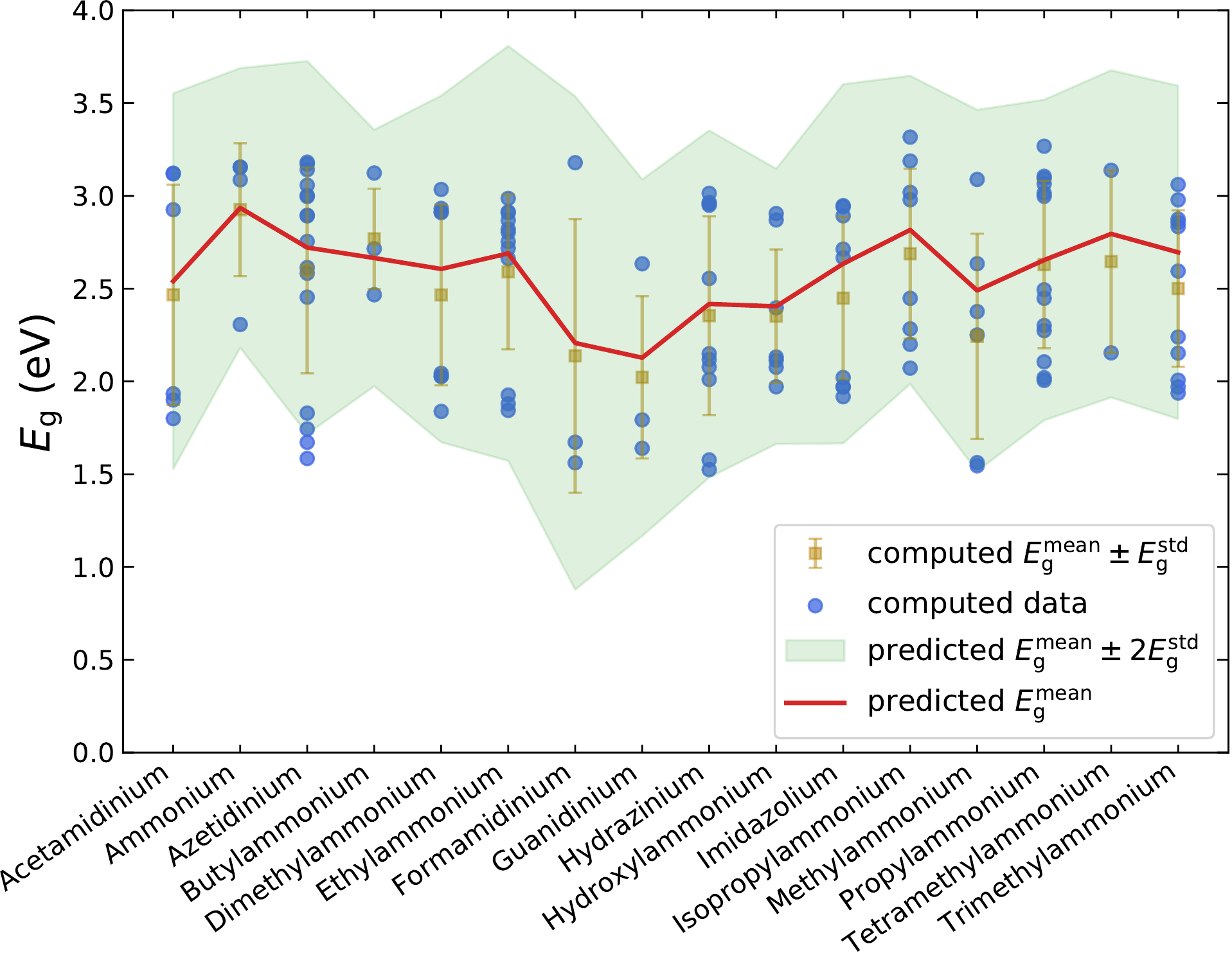}
\caption{Electronic band gap $E_{\rm g}$ (circles) computed \cite{Chiho:hybrid} for the predicted atomic structures of ASnI$_3$, 16 HOIP formulas corresponding to 16 organic cations A. For each formula, the mean and standard deviation of $E_{\rm g}$, i.e., $E_{\rm g}^{\rm mean}$ and $E_{\rm g}^{\rm std}$, are given by dark golden squares and associated errorbars. Predicted $E_{\rm g}^{\rm mean}$ is given in red while the shaded area indicates the interval of predicted $E_{\rm g}^{\rm mean} \pm 2E_{\rm g}^{\rm std}$ of the predictions using the probabilistic model developed in this work.}\label{fig:data}
\end{figure}

In fact, assuming a representative for each chemical formula is a quick response to a more general and important question of how to explore the materials space efficiently. However, the selected prototype structure is {\it not} the ground state of many ABX$_3$ formulas \cite{Huan:perovskites, Chiho:hybrid, flores2018emergence}, introducing some inevitable uncertainty that will be subsequently elaborated. The development of modern atomic structure prediction methods \cite{OganovBook} can provide a more reliable but expensive answer, using which bigger and extremely more diverse datasets were created \cite{Chiho:hybrid, Huan:Data, TuocSnS, Huan:Mg-Si}. One of them, the targeted HOIPs dataset of this work \cite{Chiho:hybrid}, contains 1,346 atomic structures of 192 ABX$_3$ formulas and some associated properties computed using density functional theory (DFT) \cite{DFT1, DFT2}. These atomic structures exhibit a wide range of different structural motifs, including not only 3D but also lower-dimensional (2D, 1D, and 0D) HOIPs \cite{Chiho:hybrid,Huan:perovskites, HuanData2021}. This diversity is unmistakably translated into a remarkable diversity in physical properties \cite{HuanData2021}, e.g., $E_{\rm g}$ is low ($\sim 1.5-2.0$ eV) with 3D HOIPs while for lower dimensions, $E_{\rm g}$ becomes higher ($\gtrsim 2.5$ eV) \cite{Mosconi,Chiho:hybrid,Huan:perovskites}. On the other hand, there is essentially no correlation between the atomic structural motifs (and the band gap) with their thermodynamic stability, as discussed and confirmed in previous structure prediction works \cite{Huan:Zn}.

Fig. \ref{fig:data} shows that each formula is associated with not a single value, but a distribution of $E_{\rm g}$ computed for a number of atomic structures of the same formula. In other words, these structures are indistinguishable when being observed by their chemical formula, but they are significantly different when atomic details are considered. This implies that estimating $E_{\rm g}$ for a chemical formula without any atomic details will suffer from an irreducible ``uncertainty'' that is closely associated with its complex energy landscape. This ``uncertainty'' is fundamentally rooted from the standpoint selected by the observer, who relies only on the information encoded in the chemical formula when making predictions. In the language of uncertainty quantification, this is an aleatoric uncertainty \cite{mcclarren2018uncertainty, abdar2020review}.

Data uncertainty of this nature is very common in materials science. The outcome of the same experimental measurement generally fluctuates each time it is repeated \cite{white2008basics, chen2021high}. Computational methods always involve some levels of approximations \cite{DFT1,DFT2}, leading to certain implicit errors \cite{wang2020uncertainty}. Such uncertainty can not be reduced or eliminated by having more data, making it different from epistemic uncertainty, whose nature is the sparsity of data \cite{mcclarren2018uncertainty, abdar2020review}, and can somehow be captured by methods like Gaussian process regression (GPR) \cite{GPRBook, GPR95}. Traditionally, when having such data uncertainty, a representative, i.e., the mean, median, minimum value, or maximum value of the available data, was used \cite{jha2019impact, sahu2021informatics}. Overall, the management of materials science data uncertainty remains in an early stage and should be promoted.

We take this opportunity to address a general problem of how to quantify the prediction ``uncertainty'' when multiple outcomes of identical observations exist. For demonstration purpose, we will focus on the HOIPs $E_{\rm g}$ dataset \cite{Chiho:hybrid}, one of many properties needed for a solar absorber, e.g., efficiency, stability, and toxicity. In particular, we designed some protocols for learning some key parameters, i.e., the mean value $E^{\rm mean}_{\rm g}$ and/or the standard deviation $E^{\rm std}_{\rm g}$, or the probability distribution of $E_{\rm g}$ directly \cite{ghahramani2015probabilistic}. We found that probabilistic deep learning (PDL) \cite{pml1Book, tfp, duerr2020probabilistic}, which treats each data entry as a distribution rather than a single number, is a robust and suitable approach. We then used the developed probabilistic model to screen over 1,284 new possible ABX$_3$ formulas, identifying those with suitable $E_{\rm g}$, the conclusion that was validated against new computations. This work is closed by a discussion in Sec. \ref{sec:remark}, extensively elaborating the possible applications of probabilistic deep learning. Given that the inevitable uncertainty in materials data should be addressed properly \cite{wang2020uncertainty,mcclarren2018uncertainty, abdar2020review}, PDL is powerful, generic, and can straightforwardly be used for numerous problems of this nature in materials informatics \cite{jha2019impact, sahu2021informatics}.

\section{Methodologies}
\subsection{Machine learning}
\subsubsection{Data and ``uncertainty''}\label{sec:data}
The targeted dataset contains 1,346 crystal structures of 192 formulas assembled from 16 organic cations A, 3 group-14 elements (Ge, Sn, and Pb) for cation B and 4 halides (F, Cl, Br, and I) for anion X \cite{Chiho:hybrid}. These organic cations (10 ammonium, 2 amidinium, and 4 others), are made up from C, N, H, and O. For each formula, low-energy structures were predicted using the minima-hopping method \cite{Goedecker:MHM, Amsler:MHM}, and then several properties were computed using DFT (details on the minima-hopping structure prediction method are given in Sec. \ref{sec:comput}). Because the structure search was completely unconstrained without any assumptions, this dataset is extremely diverse in terms of geometry, containing numerous 3D, 2D, 1D, and 0D structural motifs \cite{Chiho:hybrid}. Fig. \ref{fig:data} shows the band gap $E_{\rm g}$ computed for 122 atomic structures of 16 formulas ASnI$_3$. For each of them, multiple low-energy structures lead to a distribution of $E_{\rm g}$, which somehow connects to its actual energy landscape and how thorough it was explored during the structure searches.

\begin{table*}[t]
\caption{Three datasets (${\cal S}_1$, ${\cal S}_2$, and ${\cal S}_3$) and five models (${\cal M}_1$, ${\cal M}_2$, ${\cal M}_3$, ${\cal M}_4$, and ${\cal M}_5$) developed in this work. For neural network (NN)-based models, the number of hidden layers, the number of nodes per layer, and the choice of activation function are given.}\label{table:model}
\begin{center}
\begin{tabular}{c c c c c c c}
\hline
\hline
	Dataset & Content & Size & Model &  Algorithm & Library& Parameters\\
\hline
	 ${\cal S}_1$  & $E_{\rm g}^{\rm mean}$                           &192&${\cal M}_1$ & GPR  & SKL &N/A\\
	 ${\cal S}_1$  & $E_{\rm g}^{\rm mean}$                           &192&${\cal M}_2$ & NN   & TF & 1 layer, 5 nodes, selu\\
	 ${\cal S}_2^{\rm a}$  & $E_{\rm g}^{\rm mean}$ \& $E_{\rm g}^{\rm std}$  &192&${\cal M}_3$ & NN   & TF & 2 layers, 5 nodes, selu \\
	 ${\cal S}_2^{\rm b}$  & $E_{\rm g}^{\rm mean}$ \& $E_{\rm g}^{\rm std}$  &384&${\cal M}_4$ & NN   & TF & 2 layers, 4 nodes, tanh \\
	 ${\cal S}_3$  & $E_{\rm g}$                                      &1346&${\cal M}_5$ & PDL  & TFP & 2 layers, 5 nodes, elu  \\
\hline
\end{tabular}
\end{center}
\end{table*}

A screening over numerous formulas of ABX$_3$ can not rely on the atomic details obtained from extremely heavy calculations like DFT-based structure predictions. Therefore, one needs to evaluate the properties of interest (which is $E_{\rm g}$ in our case) solely from its formula, given the identity of A, B, and X. At this level of information, the targeted dataset has only 192 distinct entries, each of them contains a chemical formula and a distribution of $E_{\rm g}$ whose mean is $E_{\rm g}^{\rm mean}$ and standard deviation is $E_{\rm g}^{\rm std}$. For our learning purposes (see Sec. \ref{sec:learning}), we compiled three datasets, each of them encodes a level of details of the distribution. The first one, denoted by ${\cal S}_1$, contains 192 values of $E_{\rm g}^{\rm mean}$ while the second one, i.e., ${\cal S}_2$, contains 192 pairs of $E_{\rm g}^{\rm mean}$ and $E_{\rm g}^{\rm std}$. The last dataset, named ${\cal S}_3$, contains 192 distributions of $E_{\rm g}$. From the technical point of view, the whole original dataset of 1,346 entries were actually used on behalf of ${\cal S}_3$ and the probabilistic learning scheme will recognize the underlying distributions. A summary of these datasets is given in Table \ref{table:model}. All the datasets, i.e., ${\cal S}_1$, ${\cal S}_2$ (this dataset has in fact two versions, ${\cal S}_2^{\rm a}$ used for ${\cal M}_3$ and ${\cal S}_2^{\rm b}$ used for ${\cal M}_4$), and ${\cal S}_3$ are available at \texttt{www.matsml.org} and in the Supplemental Material.

\subsubsection{Features}\label{sec:fingerprint}
Our data were featurized using Matminer \cite{WARD201860}, a package that offers a rich variety of materials features at multiple levels of details, e.g., compositions and atomic structures. Because our screening will be performed over a set of new HOIP formulas, only the features that can be obtained from a chemical formula were selected. One deficiency of this selection is that the chemical composition used by Matminer is not enough to distinguish some organic cations. For examples, both ethylammonium and dimethylammonium are represented by C$_2$NH$_8$ in Matminer, although their chemical structures are different, i.e., CH$_3$-CH$_2$-NH$_3$ for the former and CH$_3$-NH$_2$-CH$_3$ for the latter. In other words, the concept of ``composition'' in Matminer is not entirely identical with the concept of ``formula'' needed for this work, in which A, B, and X must be unambiguously identified. Therefore, we augmented the Matminer composition features by a set of atomic-motif based features introduced in Ref. \onlinecite{Huan:design} that can capture such delicate differences. Within the development phase of these models, optimal sets of features were determined using the recursive feature elimination algorithm as implemented in Scikit-Learn (SKL) library \cite{scikit-learn}. 

\begin{figure}[b]
\centering
\includegraphics[width=1.0\linewidth]{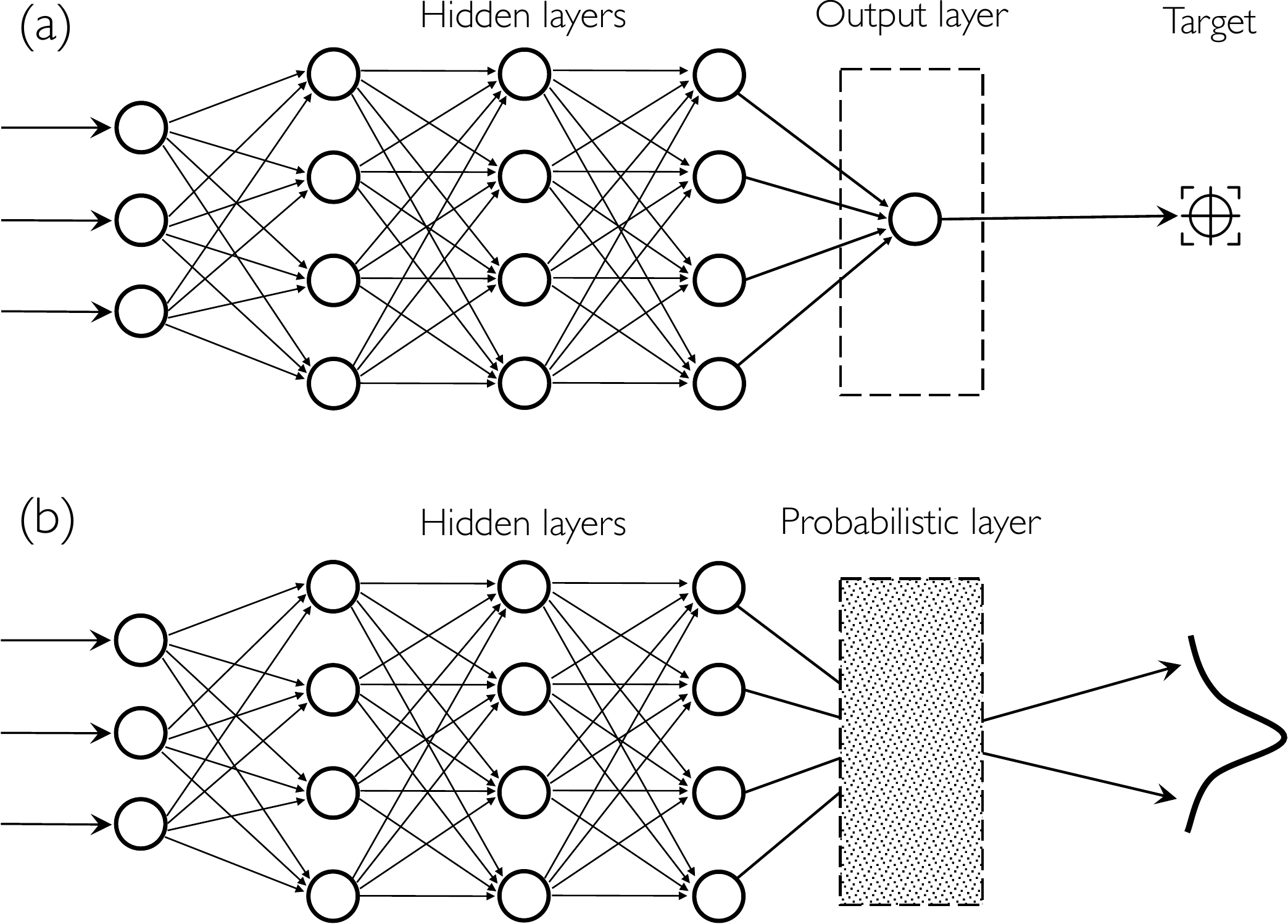}
	\caption{(a) A feedforward fully connected NN with 1 output (as used for ${\cal M}_2$ and ${\cal M}_4$ with some different parameters) and (b) a probabilistic NN (as used for ${\cal M}_5$ with some different parameters). The NN used for ${\cal M}_3$ is just a 2-outputs version of (a).}\label{fig:pnn}
\end{figure}

\subsubsection{Learning algorithms and technical details}\label{sec:learning}
A major task of this work is to learn a dataset with inevitable (intrinsic) uncertainty. Towards this goal, a hierarchy of 3 datasets encoding 3 levels of information, were prepared. Learning small datasets with single target like ${\cal S}_1$ is typical in materials informatics, and regression methods like GPR \cite{GPRBook, GPR95} are highly preferable \cite{Huan:design, Huan:FF19,doan2020machine} because they are explicitly similarity based and intuitive. Herein, we used GPR as implemented in Scikit-Learn and a simple feedforward fully connected neural network (NN) as implemented in TensorFlow (TF) and Keras to develop two baseline models, namely ${\cal M}_1$ and ${\cal M}_2$, on ${\cal S}_1$. Dataset ${\cal S}_2$ is also small, but because both $E_{\rm g}^{\rm mean}$ and $E_{\rm g}^{\rm std}$ should be learned, NN is more suitable \cite{kuenneth2021polymer}. This dataset has two versions, ${\cal S}_2^{\rm a}$ with 192 entries and ${\cal S}_2^{\rm b}$ with 384 entries. In ${\cal S}_2^{\rm a}$, $E_{\rm g}^{\rm mean}$ and $E_{\rm g}^{\rm std}$ are separate while in ${\cal S}_2^{\rm b}$, they are stacked together  by using an additional vector, whose value is either (1,0) or (0,1). Two multi-task learning models, namely ${\cal M}_3$ and ${\cal M}_4$ were developed by learning ${\cal S}_2^{\rm a}$ and ${\cal S}_2^{\rm b}$, respectively. For the aforementioned four models, root-mean-square error ($\delta^{\rm rmse}$) was used as the loss function.

The uncertainty of $E_{\rm g}$ distribution buried in ${\cal S}_3$ can be approached directly using a probabilistic deep learning approach as supported by the TensorFlow Probability (TFP) library \cite{tfp, duerr2020probabilistic}. A typical probabilistic NN contains a {\it probabilistic layer} stacked with the last hidden layer of a regular NN, treating the output as a probability distribution, but not a single value. Because $\delta^{\rm rmse}$ does not weight different points in a distribution properly, the negative log-likelihood \cite{duerr2020probabilistic} was used as the loss function in the development of ${\cal M}_5$. This model does not require $E_{\rm g}^{\rm mean}$ and $E_{\rm g}^{\rm std}$; it instead accepts the whole dataset of 1,346 fingerprinted data entries, although there are only 192 of them are distinct in terms of (composition) features. An illustration of a probabilistic NN model is given in Fig. \ref{fig:pnn} while technical details of this approach can be found in Refs. \onlinecite{tfp, duerr2020probabilistic}.

Among 5 models, ${\cal M}_1$ and ${\cal M}_2$ do not capture the data uncertainty. At the next level, ${\cal M}_3$ and ${\cal M}_4$ were designed to simultaneously learn both the mean and the standard deviation of the data. Finally, ${\cal M}_5$ targets at the possible distribution buried in the data, and thus does not need these parameters to be determined beforehand. This is an important advantage because in practice, categorizing data and computing the distribution parameters are often challenging themselves. For all five models, which are summarized in Table \ref{table:model}, overfitting and underfitting were meticulously evaluated and prevented by five-fold cross validation and searching for optimal model parameters. Within this procedure, the test data is completely unseen to the model, which was developed by repeatedly training and validating trial models in an inner loop of 5 cross-validation folds, and the best one was then selected.

\begin{figure}[t]
\centering
\includegraphics[width=1.0\linewidth]{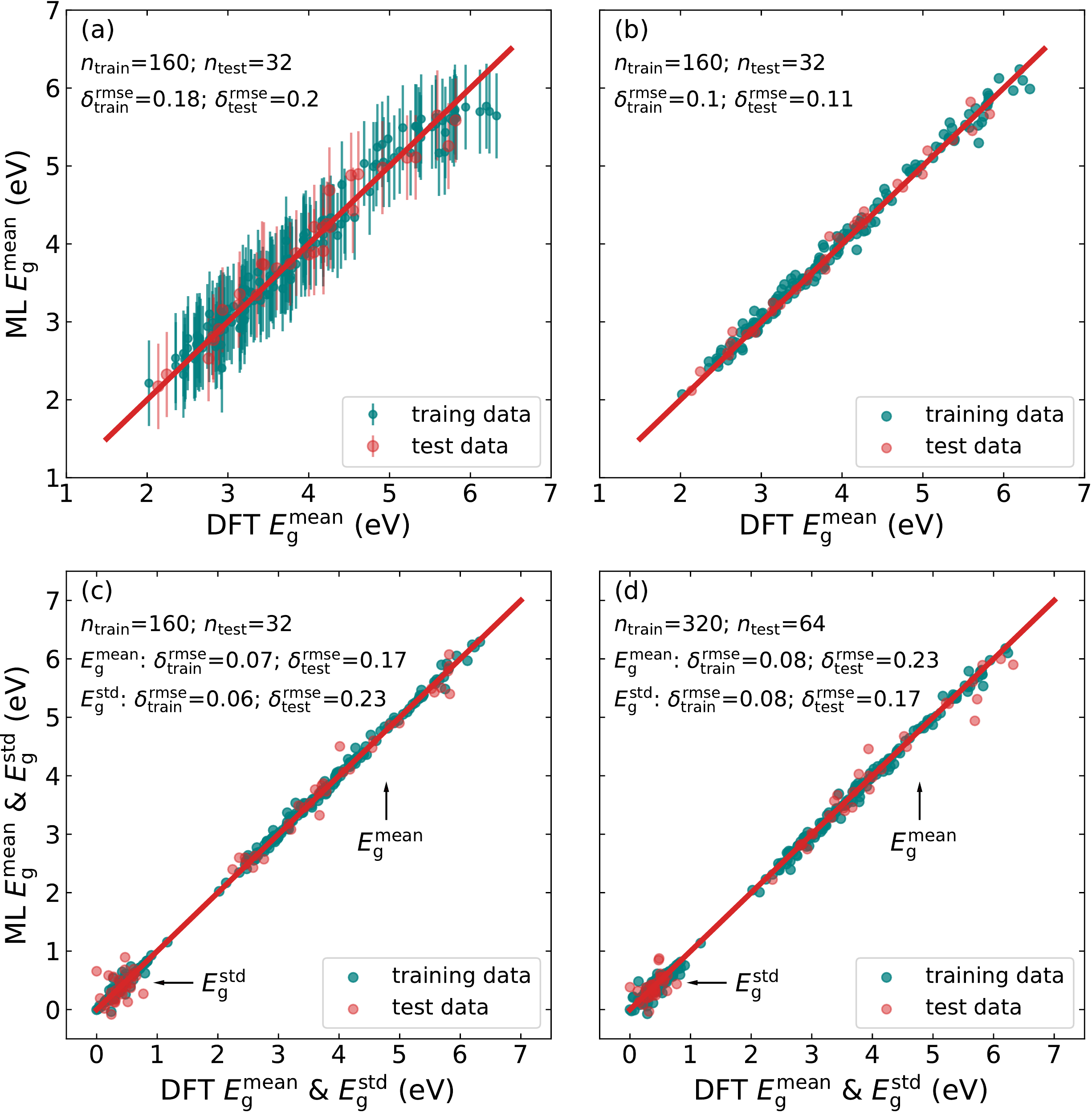}
\caption{Training/test predictions of (a) ${\cal M}_1$, (b) ${\cal M}_2$, (c) ${\cal M}_3$, and (d) ${\cal M}_4$, four models developed by learning ${\cal S}_1$ and ${\cal S}_2$. Learning curves of these models can be found in Supplemental Material \cite{supplement}.}\label{fig:model}
\end{figure}

In this work, the development of ${\cal M}_5$ aims specifically at aleatoric uncertainty, an intrinsic attribute of the computed band gap dataset of HOIPs observed from the chemical formula standpoint. Different from aleatoric uncertainty, epistemic uncertainty has its roots in the sparsity of the data in a given domain, and thus having more data in this domain could reduce the uncertainty. A classic method to handle this kind of uncertainty is Gaussian process regression, a Bayesian learning technique \cite{GPRBook, GPR95}. In fact, the errorbars shown in Fig. \ref{fig:model}(a) are a measure of the epistemic uncertainty quantified by ${\cal M}_1$, a GPR-based model. TensorFlow Probability does support GPR by using a variational Gaussian process layer \cite{tfp, duerr2020probabilistic}, thus both aleatoric and epistemic uncertainties can be handled with this package, and this could be an interesting future topic within the domain of materials informatics. The Jupyter Notebook detailing the development of 5 ML models is also available in the Supplemental Material \cite{supplement} and as an example (\texttt{ex4\_hoips}) of \texttt{matsML}, a ML toolkit for materials science, available at \texttt{https://github.com/huantd/matsml.git}.

\subsection{First-principles computations}\label{sec:comput}
The main goal of our DFT calculations is to validate $E_{\rm g}$ predicted for new HOIP formulas identified during the screening process. Therefore, the numerical scheme used for developing the original dataset \cite{Chiho:hybrid} was replicated. In particular, Vienna \textit{Ab initio} Simulation Package ({\sc vasp}) \cite{vasp3,vasp4} was used, employing a basis set of plane waves with kinetic energy up to 400 eV to represent the Kohn-Sham orbitals. The van der Waals (non-bonding) dispersion interactions between the organic cations A and the inorganic BX$_3$ frameworks were estimated using the non-local density functional vdW-DF2 \cite{vdW-DF2} while refitted Perdew-Wang 86, the generalized gradient approximation functional associated with vdW-DF2, was used to estimate the exchange-correlation (XC) energy. The Brillouin zone was sampled by a $\Gamma$-centered equispaced Monkhorst-Pack {\bf k}-point mesh with the spacing of $0.2$\AA$^{-1}$. Convergence in optimizing the structures was assumed when the atomistic forces become less than 0.01 eV/\AA. The reported electronic band gap was computed on top of the structures optimized, using the Heyd-Scuseria-Ernzerhof XC functional \cite{HSE}. The spin-orbit coupling was not included in our calculations because it was also not included in the preparation of the original dataset \cite{Chiho:hybrid}.

When new HOIP formulas were identified, their low-energy atomic structures were predicted using the minima-hopping method \cite{Goedecker:MHM, Amsler:MHM, MHM:OrganovBookChapter}, the same approach used to create the original dataset \cite{Chiho:hybrid}. For each formula, its energy landscape was constructed and explored at the DFT level of computations. Because MHM allows for unconstrained searches with strong bias toward the low-energy domains, it is powerful in identifying low-energy structures of solids, specifically those with exotic/unusual structural motifs at the atomic level \cite{Huan:Zn, Huan:hafnia, Huan:perovskites, Huan:Alanates, Chiho:hybrid}. Methods of this kind are computationally expensive and should be used for a limited set of candidates selected from large-scale screening over numerous formulas. 

\begin{figure}[t]
\centering
\includegraphics[width=1.0\linewidth]{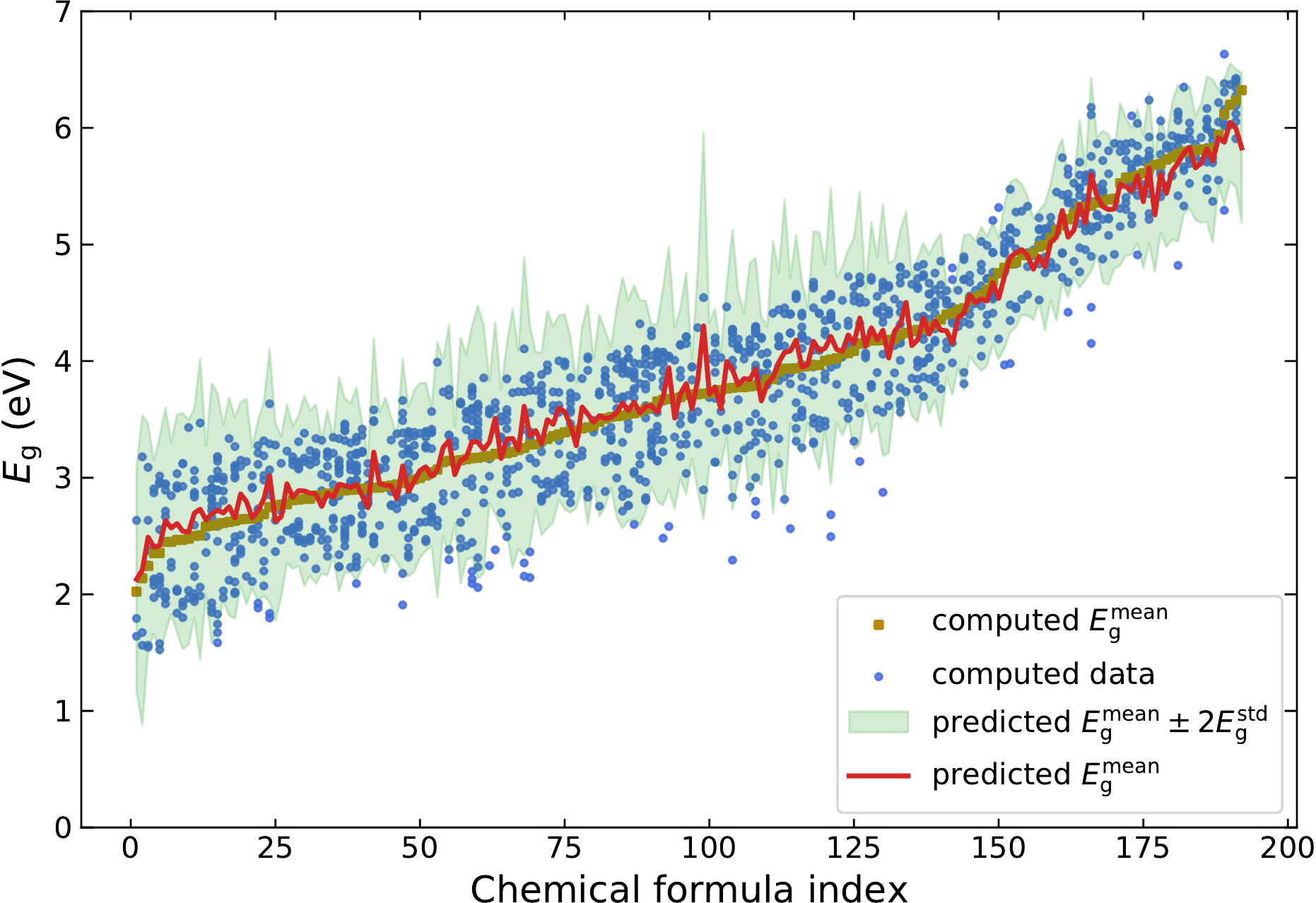}
\caption{Computed band gap $E_{\rm g}$ (dark-blue circles) in dataset ${\cal S}_3$, which includes 1,346 entries arranged in 192 chemical formulas and sorted in the ascending order of $E_{\rm g}^{\rm mean}$ (dark-golden squares). Predictions of $E_{\rm g}^{\rm mean}$ and the interval $E_{\rm g}^{\rm mean} \pm 2E_{\rm g}^{\rm std}$ using model ${\cal M}_5$ are given by the red curve and the shaded area, respectively.}\label{fig:M5}
\end{figure}

\section{Machine-learning models}
As shown in Fig. \ref{fig:model} and the learning curves given in Supplemental Material \cite{supplement}, ${\cal M}_1$ and ${\cal M}_2$ perform very well in learning and predicting $E_{\rm g}^{\rm mean}$. The $\delta^{\rm rmse}$ of the training and test data is $\simeq 0.2$ eV for ${\cal M}_1$ and $\simeq 0.1$ eV for ${\cal M}_2$, significantly lower than the range of $\simeq 0.2 - 0.5$ eV of any previously reported works \cite{lu2018accelerated, gladkikh2020machine} when features at the elemental and composition levels were used. The errorbars in Fig. \ref{fig:model}(a) were predicted by GPR as a measure of the epistemic uncertainty, which should be smaller when more data are available --- within this approach, data are assumed to be the groundtrust with no uncertainty. Regarding learning algorithm, we found that NN, when being meticulously calibrated, can work well with small datasets and steadily outperforms GPR. Two multi-task models, i.e., ${\cal M}_3$ and ${\cal M}_4$, also yield similar small $\delta^{\rm rmse}$, ranging in $\simeq 0.1 - 0.2$ eV. This observation shows that if we can assume a given distribution of the target, its parameters can be learned and the obtained predictive models can be used for new cases. For all four models, the $R^2$ score of all the the training and test datasets is no less than 95\%, indicating that the material features generated by Matminer capture very well the underlying physics and chemistry of the materials while the learning procedure could efficiently unravel the targeted formula-property relationship. 

The predictions performed by the probabilistic model ${\cal M}_5$, which was trained on the entire dataset ${\cal S}_3$, are shown in Fig. \ref{fig:M5}. Compared to the approach of learning $E_{\rm g}^{\rm mean}$ and $E_{\rm g}^{\rm std}$ (either simultaneously or separately in the previous models), the probabilistic learning approach with TFP does not require a cumbersome preprocessing step in which the original data are categorized into different chemical formulas and then $E_{\rm g}^{\rm mean}$ and $E_{\rm g}^{\rm std}$ are computed. This advantage becomes important in numerous practical cases of continuous categories because some kinds of clustering methods will then be needed for the preprocessing. More importantly, the predicted distribution of $E_{\rm g}$ shown in Fig. \ref{fig:M5} does capture very well both the mean and the variation of the data. For $E_{\rm g}^{\rm mean}$, the cross-validation $\delta^{\rm rmse} = 0.14~{\rm eV}$ while for $E_{\rm g}^{\rm std}$, this error metric is 0.09 eV. Within the confidence interval of $E_{\rm g}^{\rm mean} \pm 2E_{\rm g}^{\rm std}$, more than 95\% of the original dataset was indeed captured. Fig. \ref{fig:M5} also reveals that for $E^{\rm mean}_{\rm g}\lesssim 4$ eV, the original data distribution is slightly asymmetric, i.e., the data points below $E_{\rm g}^{\rm mean}$ are distributed further down compared with those residing above $E_{\rm g}^{\rm mean}$. Therefore, the upper bound of the interval of $E_{\rm g}^{\rm mean} \pm 2E_{\rm g}^{\rm std}$, which is symmetric, seems to extend further up. On the other hand, the distributions with $E^{\rm mean}_{\rm g}\gtrsim 4$ eV are also asymmetric, but in the reverse direction. Therefore, the lower bound of $E_{\rm g}^{\rm mean} \pm 2E_{\rm g}^{\rm std}$ seems to overestimate the variation of the data below $E_{\rm g}^{\rm mean}$. This observation indicates that while ${\cal M}_5$ captures pretty well the data distribution (or the uncertainty), ${\cal M}_3$ and ${\cal M}_4$ will not be at the level of ${\cal M}_5$ because the first two moments, $E_{\rm g}^{\rm mean}$ and $E_{\rm g}^{\rm std}$, are not enough to represent arbitrary (asymmetric) distributions. Overall, we believe that probabilistic learning is robust and versatile, and specifically, model ${\cal M}_5$ is reliable and suitable for the HOIPs space exploration discussed in the subsequent part of this work.

\begin{figure}[t]
\centering
\includegraphics[width=1.0\linewidth]{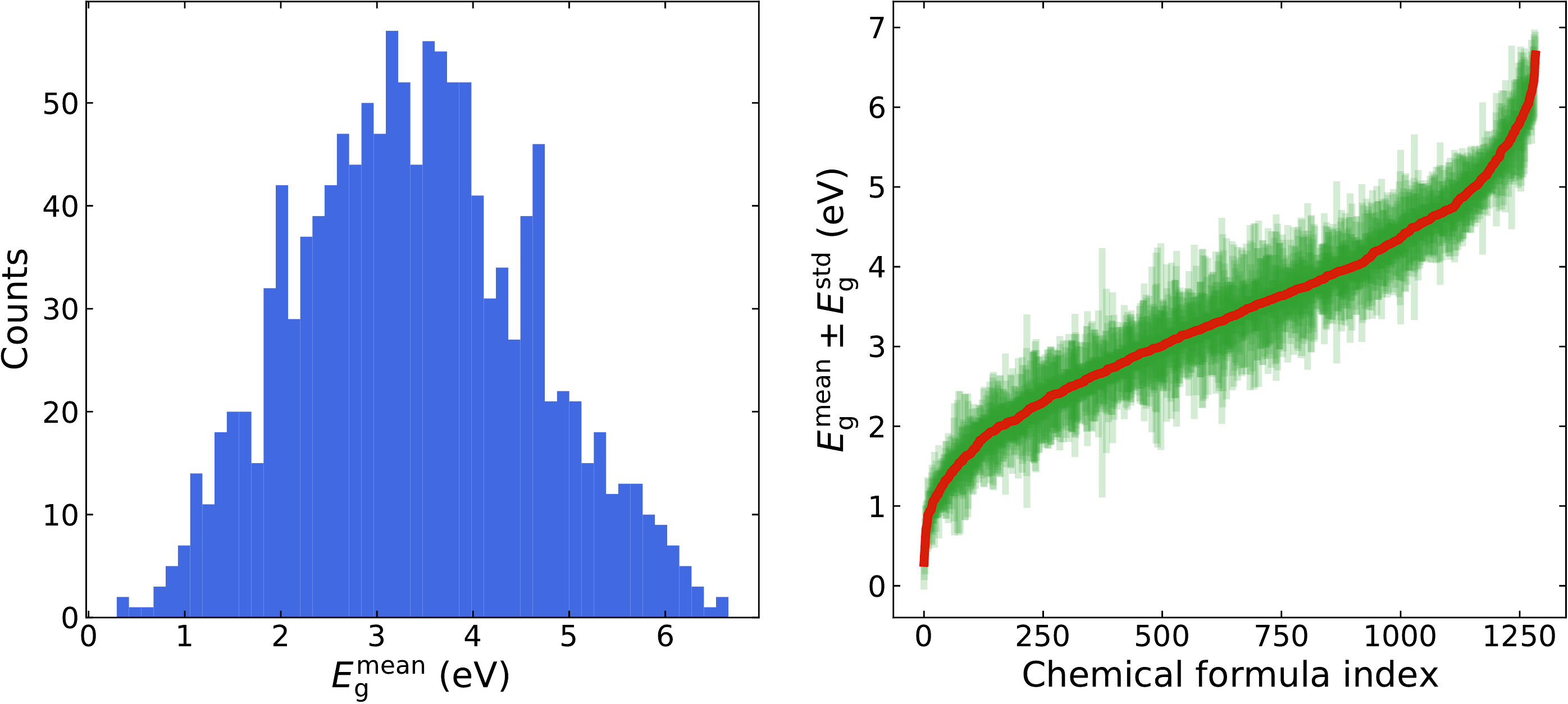}
	\caption{(a) Predicted $E_{\rm g}^{\rm mean}$ distribution and (b) sorted predictions of $E_{\rm g}^{\rm mean}$ (red curve) and $E_{\rm g}^{\rm std}$ (light green errorbars). Results obtained by using the probabilistic model ${\cal M}_5$ on the screening space of 1,284 HOIP formulas.}\label{fig:disc}
\end{figure}

\section{Machine-learning assisted exploration of HOIPs}
\subsection{Screening space}
For the purpose of uncovering new HOIPs, a new dataset of 107 monovalent, positively charged (ammonium) organic cations were obtained from small molecules synthesized and reported in the literature. We first collected the molecules that (1) have no more than 13 atoms, (2) contain at least C, H, and N, and possibly O, as covered by 16 organic cations in the learning dataset \cite{Chiho:hybrid}, and (3) have at least one N atom with a lone pair of electrons. Then, a hydrogen atom was added to create a covalent bond with this N atom, making each molecule an ammonium cation. For examples, the famous methylammonium and the trimethylammonium \cite{Chiho:hybrid} cations can be obtained by using this procedure on the methylamine, i.e., CH$_3$NH$_2$, and trimethylamine, i.e., N(CH$_3$)$_3$, molecules, respectively. In fact, this procedure mimics the reactions between these amines and some acids, e.g., hydrochloric acid, to make respective ammonium salts \cite{rawn2018organic}. Combining these new organic cations with the 3 group-14 elements and the 4 halides, a total of 1,284 new chemical formulas of HOIPs were obtained to define the screening space. Although we limited ourselves in the cations that are closely related to the dataset we used to develop the ML models, our dataset of 107 small ammonium cations, which are available in Supplemental Material \cite{supplement}, is larger than all the existing counterparts. Moreover, we note that the technical approach described in this work can be used for any set of organic cations.

\begin{figure}[t]
\centering
\includegraphics[width=1.0\linewidth]{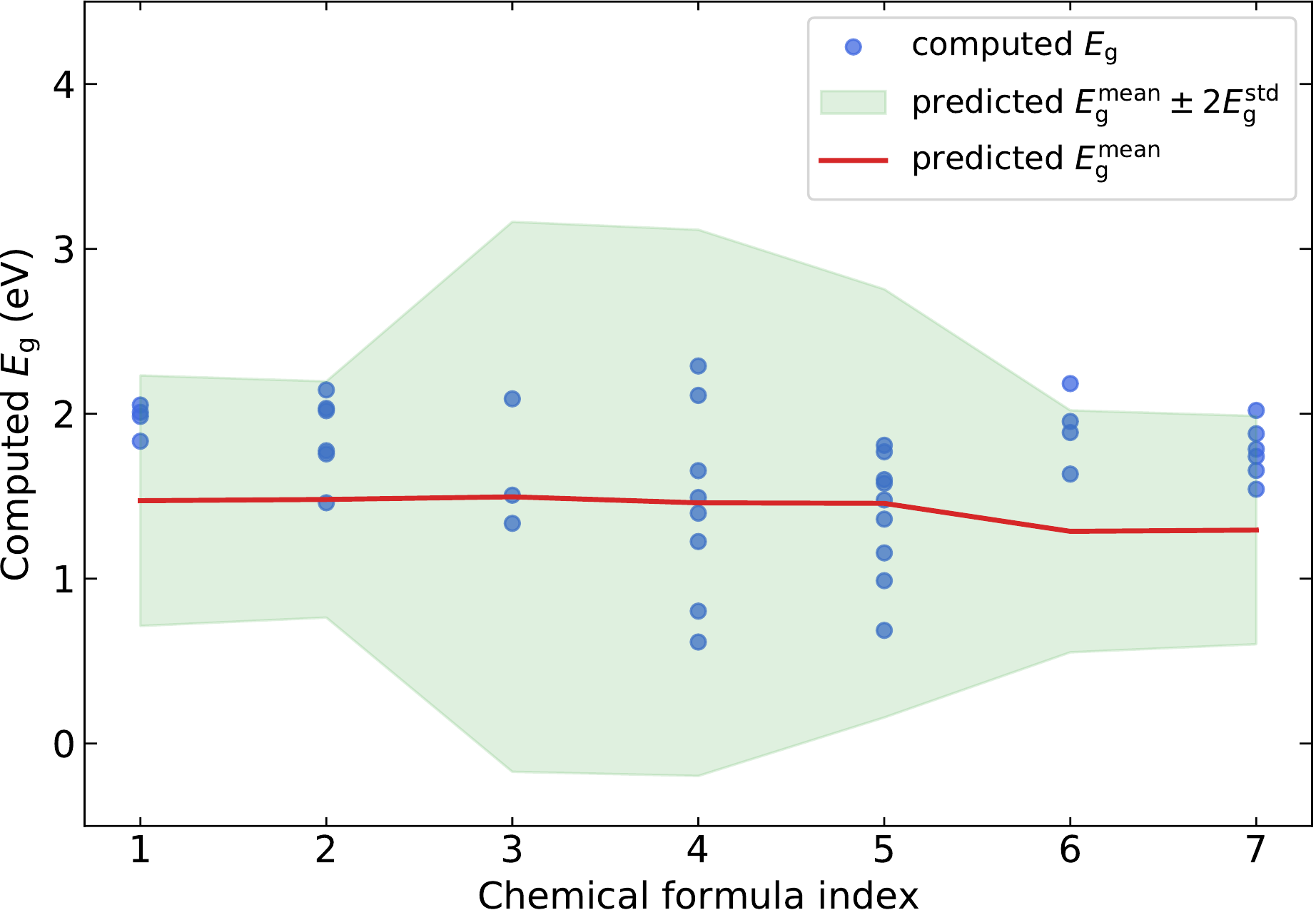}
	\caption{Predicted $E_{\rm g}^{\rm mean}$ (red line), the interval of predicted $E_{\rm g}^{\rm mean} \pm 2E_{\rm g}^{\rm std}$, (shaded area) and the computed $E_{\rm gap}$ (dark blue circles) of the atomic structures predicted for 7 selected candidates. Details on these candidates, including the chemical formula index, are given in Table \ref{table:disc}.}\label{fig:validation}
\end{figure}

\subsection{Candidates with targeted $E_{\rm g}$}
The dataset of 1,284 HOIP formulas was featurized using the approach described in Sec. \ref{sec:fingerprint} and predictions were made using the probabilistic models ${\cal M}_5$ we developed. A summary of $E_{\rm g}^{\rm mean}$ and $E_{\rm g}^{\rm std}$, the parameters of the predicted distribution of $E_{\rm g}$, is shown in Fig. \ref{fig:disc}. Within the screening space, $E_{\rm g}^{\rm mean}$ ranges from $\simeq 0.5$ eV to $\simeq 6.5$ eV, and $E_{\rm g}^{\rm std}$ provides a confidence interval for the predictions. While this range of $E_{\rm g}$ may correspond to various technology applications, we focused on the ideal window of solar cell applications, i.e., $1.25~{\rm eV} \leq E^{\rm mean}_{\rm g} \leq 1.50~{\rm eV}$. Within this range, 34 new HOIP formulas were identified and given in the Supplemental Material \cite{supplement}. Seven of them, listed in Table \ref{table:disc}, were selected for the validation. Within this step, low-energy atomic structures were predicted by the minima-hopping method, and their $E_{\rm g}$ was computed using the numerical schemes described in Sec. \ref{sec:comput}. 

\begin{table}[t]
	\caption{Seven HOIP formulas with $E_{\rm g}^{\rm mean}$ predicted to be in the range of $1.25~{\rm eV} - 1.50~{\rm eV}$ and selected for validation. For each of them, details on the the cation A, the cation B, the anion X, the predicted $E_{\rm g}^{\rm mean}$ (eV) and $E_{\rm g}^{\rm std}$ (eV), and a list of $E_{\rm g}$ (eV) computed for the predicted atomic structures, are provided. The index given in this Table was also used for Fig. \ref{fig:validation}. In the Cation ``A'' column, C, H, N, and O atoms are shown in dark brown, pink, cyan, and red, respectively.}\label{table:disc}
\begin{center}
	\begin{tabular}{  >{\centering\arraybackslash}m{0.15in} >{\centering\arraybackslash}m{0.55in} >{\centering\arraybackslash}m{0.25in} >{\centering\arraybackslash}m{0.25in} >{\centering\arraybackslash}m{0.3in} >{\centering\arraybackslash}m{0.3in} >{\centering\arraybackslash}m{1.2in}}
\hline
\hline
		Id & A & B & X & $E_{\rm g}^{\rm mean}$ &  $E_{\rm g}^{\rm std}$ & Computed $E_{\rm g}$\\
\hline
		1 & \includegraphics[width=0.8cm]{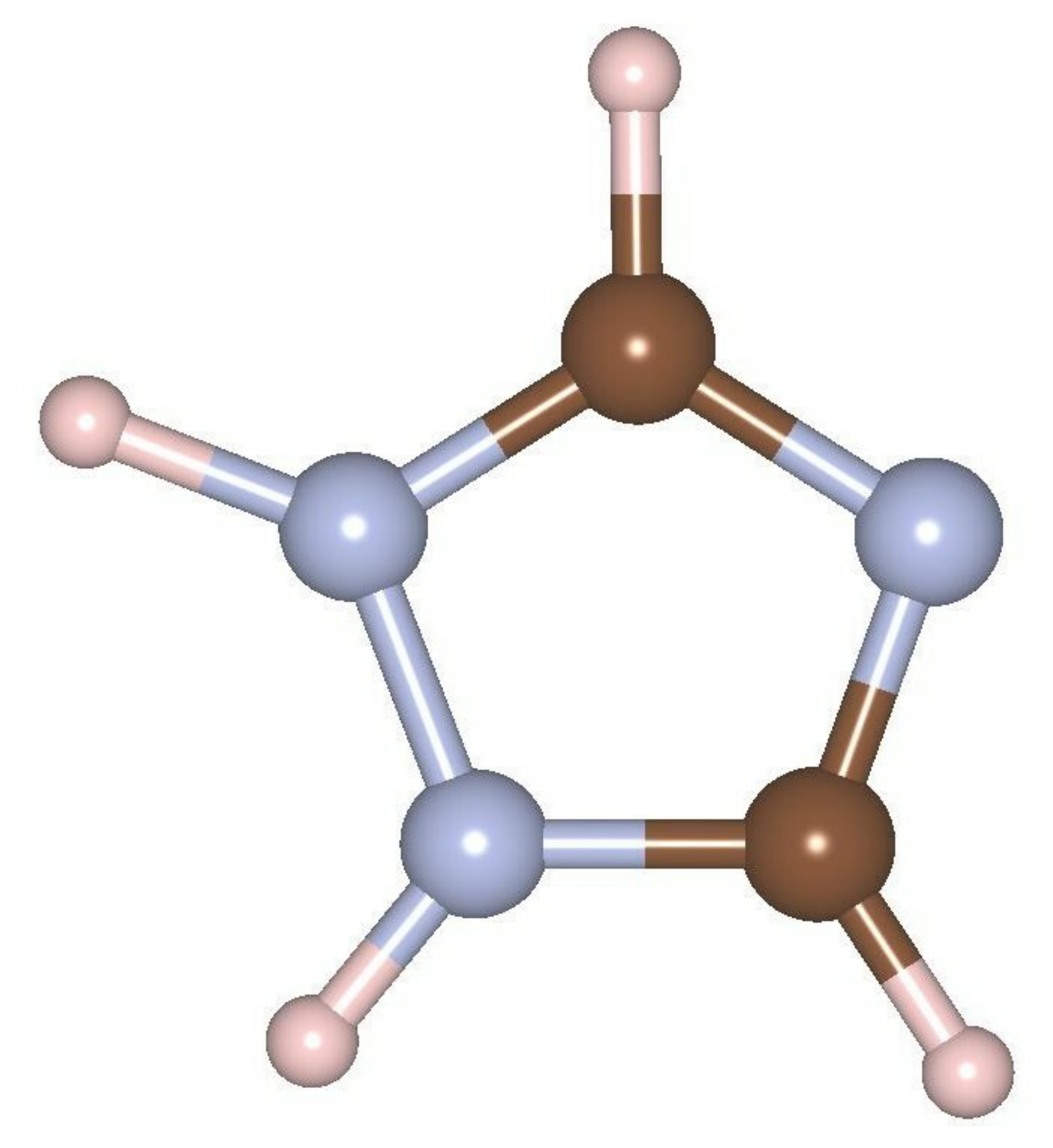} & Sn  &Br & 1.47 & 0.38& 1.83, 1.98, 2.00, 2.05 \\
		\hline
		2 & \includegraphics[width=0.7cm]{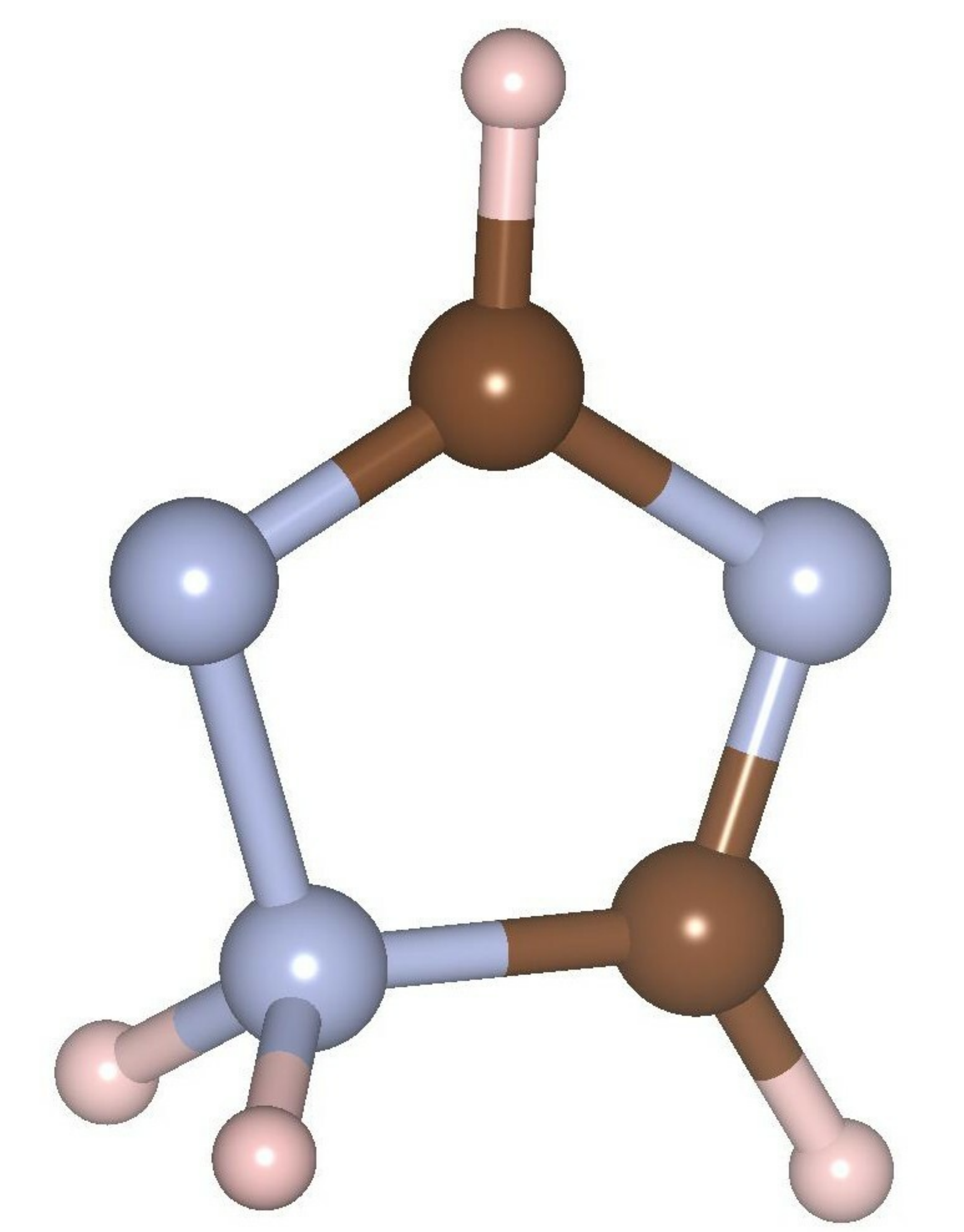} & Sn  &Br & 1.47 & 0.36 & 1.46, 1.75, 1.78, 2.01, 2.03, 2.15\\
		\hline
		3 & \includegraphics[width=1.0cm]{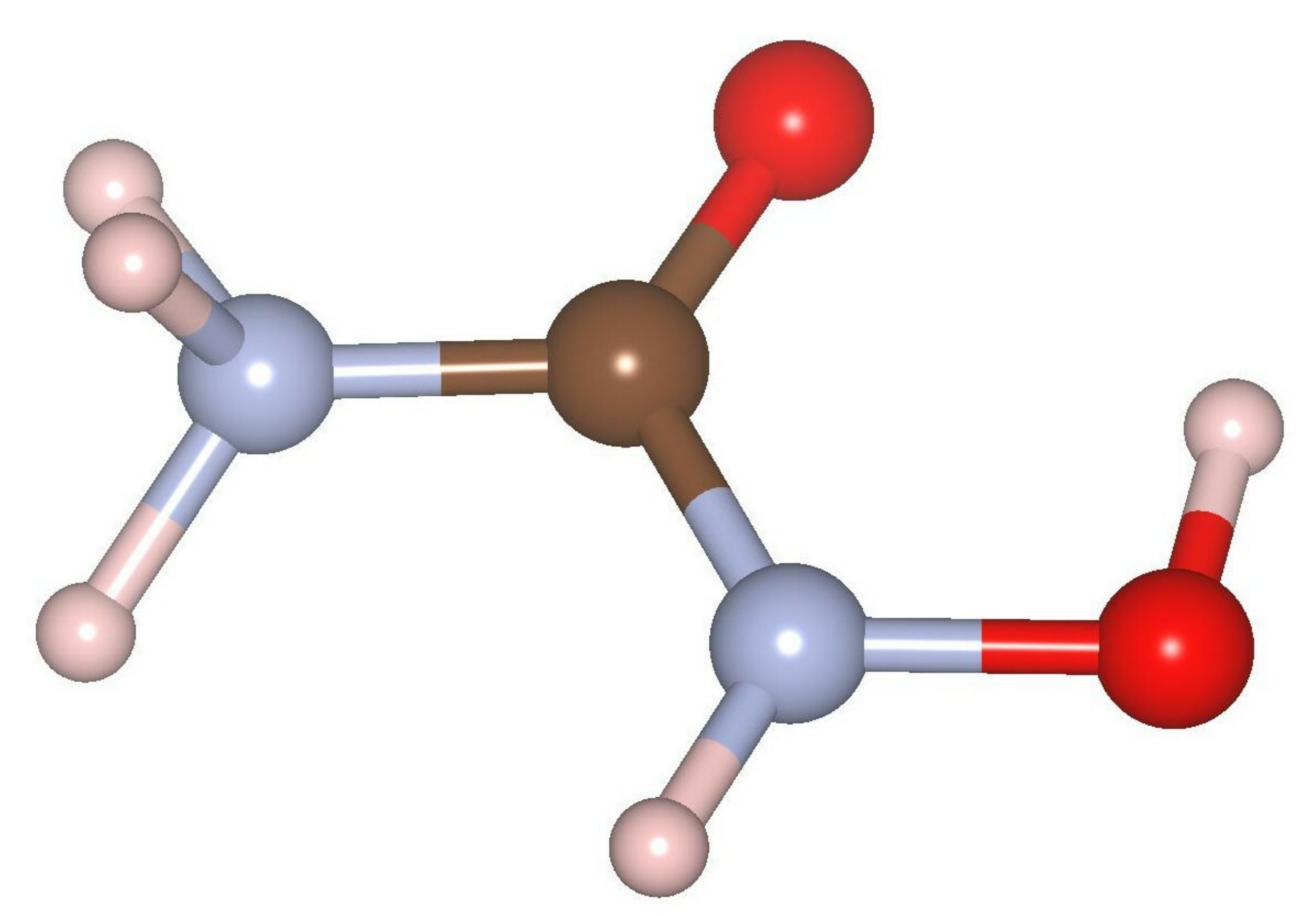} & Sn  &I & 1.50 & 0.83 & 1.34, 1.50, 2.10\\
		\hline
		4 & \includegraphics[width=1.0cm]{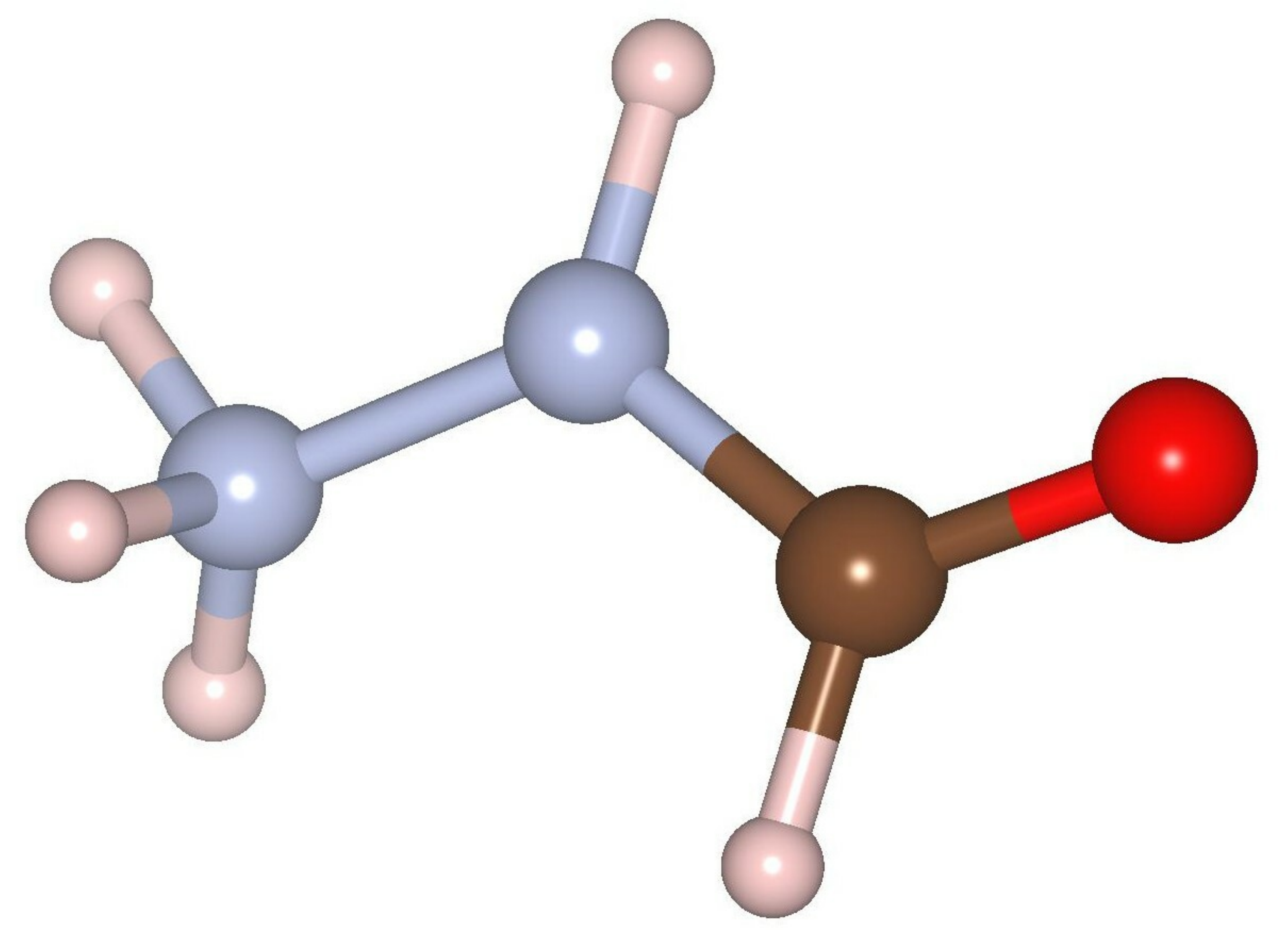} & Sn  &I & 1.46 & 0.83 & 0.62, 0.80, 1.22, 1.40, 1.49, 1.65, 2.11, 2.29\\
		\hline
		5 & \includegraphics[width=0.9cm]{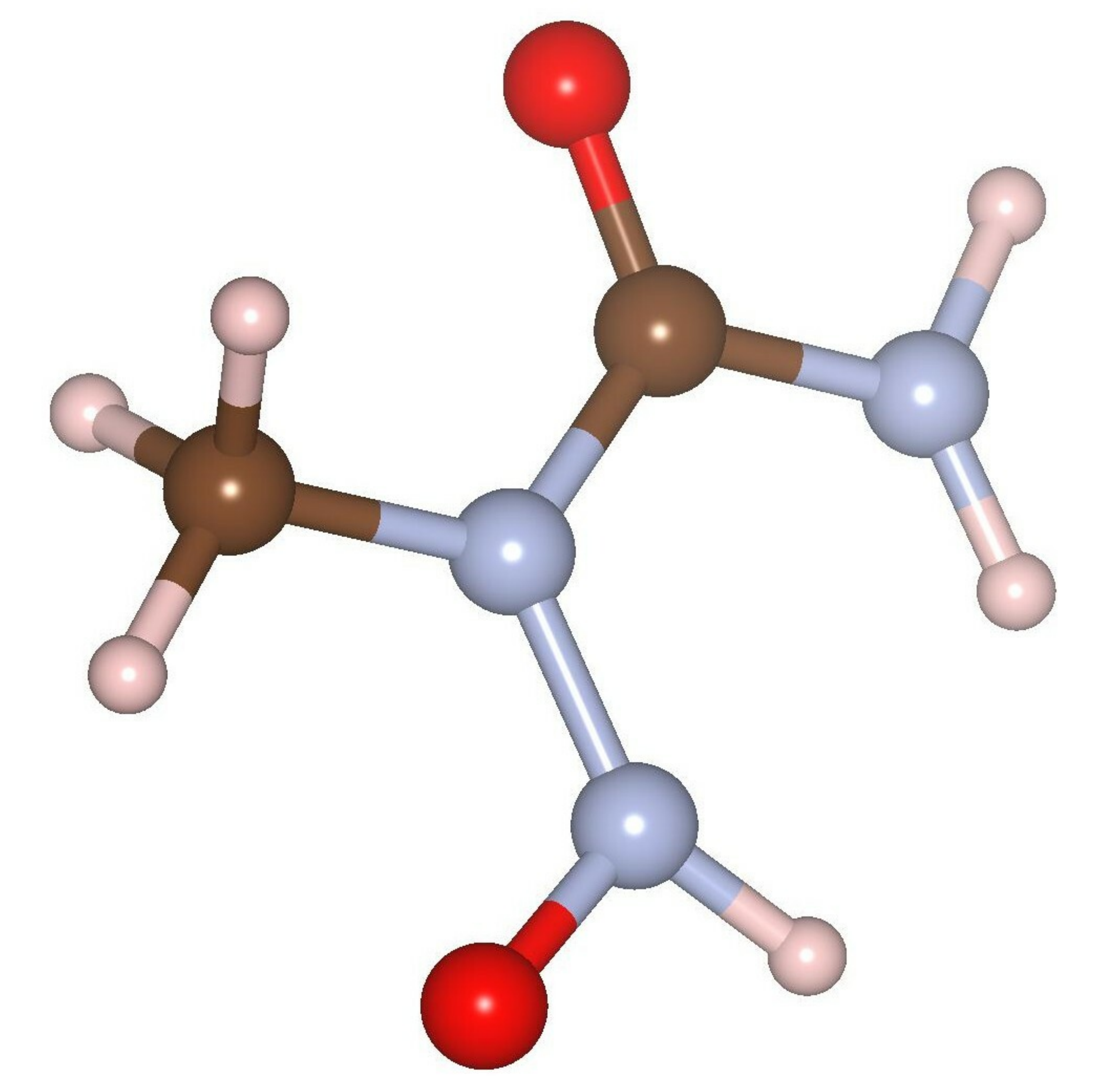} & Sn  &I & 1.46 & 0.65 & 0.69, 0.99, 1.16, 1.36, 1.48, 1.60, 1.77\\
		\hline
		6 & \includegraphics[width=0.8cm]{tb2-cat1.pdf} & Pb  &I & 1.29 & 0.37 & 1.63, 1.89, 1.95, 2.18\\
		\hline
		7 & \includegraphics[width=0.7cm]{tb2-cat2.pdf} & Pb  &I & 1.29 & 0.35 & 1.54, 1.66, 1.74, 1.79, 1.88, 2.02\\
\hline
\end{tabular}
\end{center}
\end{table}

As shown in Table \ref{table:disc} and Fig. \ref{fig:validation}, the obtained results of $E_{\rm g}$ fall very well within the 95-percent confidence interval rendered using the probabilistic model ${\cal M}_5$. Among 7 cases considered, ${\cal M}_5$ slightly underestimates $E_{\rm g}^{\rm mean}$ by about $0.3$ eV for 5 cases (with indicies 1, 2, 3, 6, and 7), slightly overestimates $E_{\rm g}^{\rm mean}$ by about $0.1$ eV for one case (5), and correctly predicts $E_{\rm g}^{\rm mean}$  for 1 case (6). For some cases, the $E_{\rm g}^{\rm std}$ seems to be slightly overestimated, probably indicating that the training data is still not large and diverse enough. This observation indicates that while ${\cal M}_5$ model can be used for the designated screening of HOIPs based on Pb, Sn, and Ge for B and F, Cl, I, and Br for X, it should be progressively improved whenever new data become available.

\section{Remarks and outlook}\label{sec:remark}
The key trademark of probabilistic deep learning is the standpoint from which realistic data with irreducible uncertainty are treated. As discussed in Sec. \ref{sec:intro}, it is very common in materials science to have multiple values of a given materials property while the nature of this divergence may never be entirely clear. Without further information, these values define an uncertainty characterized by a distribution function of which some characteristics like the mean value are used within traditional approaches \cite{jha2019impact, sahu2021informatics}. PDL offers a more appropriate and complete way to treat this kind of data, directly recognizing the distribution functions. While a specific dataset of HOIPs was used herein for a demonstration, PDL is a generic and far-reaching approach. 

Let us consider two examples for which PDL is unmistakably the best approach. After the $P6_3/mmc$ phase of CeH$_9$ was synthesized in six experimental diamond anvil cells at $\simeq 90$GPa, these chambers were compressed to nearly 200GPa and then decompressed back to $\simeq 90$GPa \cite{chen2021high}. Along these trajectories, the superconducting critical temperature $T_{\rm c}$ measured for the $P6_3/mmc$ phase of CeH$_9$ behaves significantly different, i.e., at a given pressure, there are multiple values of measured $T_{\rm c}$ of the same superconducting phase of CeH$_9$ \cite{chen2021high}. Clearly, there should be some delicate dissimilarities among these systems that warrant further studies, within which they may or may not be identified. However, given the values of $T_{\rm c}$ measured for the $P6_3/mmc$ phase of CeH$_9$ at each pressure, the best solution is to accept and process all of them. Simply put, PDL is an ideal tool for the (currently active) problem of predicting the $T_{\rm c}$ directly from material chemical formulas \cite{stanev2018machine,shipley2021high,song2021high}. 

As another example, a dataset of 1,545 entries provides some mechanical properties of 630 unique alloy compositions \cite{borg2020expanded}. For each entry, information on the microstructure and processing method may or may not be avaiable. A closer look reveals that multiple entries can be found for the same composition, microstructure, and processing method. For instance, there are 31 entries whose composition is \texttt{Co$_1$Cr$_1$Fe$_1$Mn$_1$Ni$_1$}, microstructure is \texttt{FCC}, and processing method is \texttt{WROUGHT}. There are 6 other entries of which the composition is \texttt{Al$_1$Co$_1$Cr$_1$Fe$_1$Ni$_1$}, the microstructure is \texttt{BCC}, and the processing method is \texttt{CAST}. In any cases, the mechanical properties are different accross the avaiable entries, implying that the information we have is {\it incomplete} to fully describe the mechanical properties of the alloys considered. While the mean value of these entries can be used, using PDL to treat all of these data points as parts of a distribution is clearly the most suitable approach.

\section{Conclusions}
The main results of this work are the probabilistic standpoint and approach to handle realistic materials data whose uncertainty is inevitable. The traditional approaches are deterministic in nature, i.e., a representative is selected for a set of indistinguishable observations whose outcomes may vary, forming a reduced dataset of distinguishable observations. On the other hand, PDL accepts all available observations, recognizing the entire outcome distribution for any set of indistinguishable observations. We find that PDF is a robust and complete method for directly utilizing all the the available data and quantifying the aleatoric uncertainty.

For a demonstration, we have developed a probabilistic machine learning model (${\cal M}_5$) on a dataset of 1,346 atomic structures of 192 HOIP formulas. This model was then used in a screening over 1,284 new formulas, identifying those with targeted electronic band gap $E_{\rm g}$. Within this workflow, ${\cal M}_5$ must rely only on the HOIP chemical formulas, not any atomic structures. At this level of details, each formula corresponds to multiple values of $E_{\rm g}$, and this sort of irreducible data ``uncertainty'' was properly handled within the probabilistic deep learning approach supported by TensorFlow Probability. The predictions of ${\cal M}_5$ were validated against suitable DFT computations and the list of HOIP formulas identified should be subjected to further investigations on other properties needed for specific applications. 

\section*{Acknowledgements}
Works by V.N.T. and T.D.H. were supported by Vingroup Innovation Foundation (VINIF) in project code VINIF.2019.DA03 while computational support from XSEDE through the allocation No. TG-DMR170031 is acknowledged. The authors thank V. Roshan Joseph (Georgia Institute of Technology) for useful discussions, and three anonymous reviewers whose comments have helped us to improve the paper.

%
\end{document}